\documentclass[aps,amssymb,twocolumn]{revtex4}
\usepackage{amsmath}
\usepackage{amsfonts}
\usepackage{amssymb}
\usepackage{graphics,epsfig}
\usepackage{graphicx}

\begin{document}

\title{An infinite class of exact static anisotropic spheres that break the Buchdal bound}
\author{Kayll Lake$\;$\cite{email}}
\affiliation{Department of Physics, Queen's University, Kingston, Ontario, Canada, K7L
3N6 }
\date{\today}

\begin{abstract}
An infinite class of exact static anisotropic spheres is developed. All members of the class satisfy (i) regularity (meaning no singularities), and in particular at the origin, (ii) positive but monotone decreasing energy density ($\rho(r)$), radial pressure ($p(r)$), and tangential pressure ($P(r)$), (iii) a finite value of $r=R$ such that $p(R)=0$ defining the boundary surface onto vacuum, (iv) $p \leq \rho$, and (v) $p + 2 P=3 \rho$. All standard energy conditions are satisfied except for the dominant energy condition which has an innocuous violation by the tangential stress since $\rho \leq P$ by construction. An infinite number of the solutions violate the Buchdal bound.
\end{abstract}
\maketitle

\section{Introduction}
After extensive examinations of spherically-symmetric static isotropic perfect solutions of Einstein's equations \cite{dellake}, there followed, rather quickly, a remarkable collection of solution generating techniques \cite{visser}. At the same time, interest in exact anisotropic fluid solutions of Einstein's equations has grown \cite{anisotropic}. One important reason that these solutions are of interest is that they do not have to satisfy the Buchdal bound \cite{bb}. Solution generating techniques are known for anisotropic fluids \cite{herrera}.  These require the specification of two input functions (as opposed to one in the isotropic case), typically an ``anisotropy" function for the pressure. Since \textit{any} static spherically symmetric metric is an anisotropic fluid ``solution" of Einstein's equations, it is important to clarify, at the start, what conditions are imposed on the physical parameters. Here these are:
\begin{itemize}
\item[\textbf{1.}] Regularity (meaning no singularities), and in particular at the origin,
\item[\textbf{2.}]Positive but monotone decreasing energy density ($\rho(r)$), radial pressure ($p(r)$), and tangential pressure ($P(r)$),
\item[\textbf{3.}] A finite value of $r=R$ such that $p(R)=0$ so as to define a boundary surface,
\item[\textbf{4.}] $p \leq \rho$,
\item[\textbf{5.}] $p + 2 P=3 \rho$.
\end{itemize}
All standard energy conditions are satisfied \cite{poisson} except, by construction, $\rho \leq P$ so that the tangential stress ``violates" the dominant energy condition  \cite{footnote}.
\section{Geometry}

We start with the geometry in the standard form \cite{notation}
\begin{equation}\label{metric}
    ds^2=\frac{dr^2}{1-2m(r)/r}+r^2d\Omega^2-e^{2\Phi(r)}dt^2
\end{equation}
where $d\Omega^2$ is the metric of a unit 2-sphere ($d\theta^2+sin^2\theta d\phi^2$). By way of Einstein's equations, the source of (\ref{metric}) is taken to be a comoving fluid described by the stress-energy tensor $T^{\alpha}_{\beta}=diag[p(r),P(r),P(r),-\rho(r)]$ where $p, P$ and $\rho$ are positive. From Einstein's equations we find the effective gravitational mass
\begin{equation}\label{mass}
    m=4 \pi \int _{0}^{r}\rho(x) x^2dx,
\end{equation}
the source equation for $\Phi$
\begin{equation}\label{phi}
    \Phi^{'}=\frac{m+4 \pi r^3 p}{r(r-2m)},
\end{equation}
and the generalized Tolman-Oppenheimer-Volkoff equation
\begin{equation}\label{peqn}
    P=\frac{r}{2}(p^{'}+(\rho+p) \Phi^{'})+p,
\end{equation}
where $'\equiv d/dr$. The fluid distribution is assumed to terminate at finite $r=R$ where $p(R)=0$ and to join there, by way of a boundary surface, onto a Schwarzschild vacuum of mass $m(R)\equiv M$. No further conditions need be specified for the junction.

Scalar polynomial singularities involve scalar invariants built out of the Riemann tensor. Due to the spherical symmetry assumed here, there are only four independent invariants of this type \cite{invar}. A direct calculation shows that $\rho^{'}$ and $P^{'}$ do not enter these invariants and that the Ricci invariants are regular as long as $\rho, p$ and $P$ are. However, the second Weyl invariant grows like
\begin{equation}\label{weyl}
   \left(\frac{rp^{'}+2(p-P)}{\rho-3p+4P}\right)\frac{1}{r^6}
\end{equation}
as $r \rightarrow 0$. As a consequence, we have the following necessary conditions for regularity
\begin{equation}\label{regular}
    p^{'}(0)=0, \;\;\;\;p(0)=P(0),
\end{equation}
the latter of which is already obvious from (\ref{peqn}). In terms of $\Phi$, as in the isotropic case, $\Phi$ must be a monotone increasing function of $r$ with a regular minimum at $r=0$.
\section{The Algorithm}
Rather than consider an anisotropy function, here we start with the condition
\begin{equation}\label{dominant}
p+2P=3 \rho.
\end{equation}
As shown by Andr\'{e}asson \cite{andre}, with (\ref{dominant})
\begin{equation}\label{sup}
\sup_{r>0}\frac{2m(r)}{r} \leq \frac{48}{49} .
\end{equation}
Solving (\ref{dominant}) for $m(r)$ we find
\begin{equation}
m(r) =\frac{\int \!b ( r) {e^{\int \!a ( r)
{dr}}}{dr}+\mathcal{C}}{{e^{\int \!a(r) {dr}}}} \label{newmass}
\end{equation}
where
\begin{equation}
a(r) \equiv \frac{3\Phi^{'}+2r(\Phi^{''}+(\Phi^{'})^2)}{r\Phi^{'}+4}\label{a}
\end{equation}
and
\begin{equation}
b(r) \equiv \frac{2\Phi^{'}+r(\Phi^{''}+(\Phi^{'})^2)}{r\Phi^{'}+4}\label{b}
\end{equation}
with $\mathcal{C}$ a constant. Explicit solutions are given by the functions $\Phi$ for which $m$ can be obtained without recourse to numerical integration. However, it is important to note that the detailed structure of the solutions is most easily explored via numerical methods in individual cases.

\section{An Infinite Series of Examples}
If we take
\begin{equation}
\Phi(r)=\frac{1}{2}\,N\ln( 1+{\frac
{{r}^{2}}{{\alpha}}}),\label{tolman}
\end{equation}
where $N$ is an integer $\geq 1$ and $\alpha$ is a constant $> 0$, it follows from (\ref{newmass}), (\ref{a}) and (\ref{b}) that
\begin{equation}\label{newtolman}
m(r)=c(r)d(r)
\end{equation}
where
\begin{equation}
c(r) = \frac{(r^2(N+4)+4 \alpha)^{\frac{3N-16}{2(N+4)}}}{(r^2+\alpha)^N}(r^2+\alpha)^2,\label{c}
\end{equation}
and
\begin{equation}
d(r) = \int\frac{N(r^2+\alpha)^{N-3}(r^2(N+1)+3 \alpha)r^2dr}{(r^2(N+4)+4 \alpha)^{\frac{5N-8}{2(N+4)}}}.\label{d}
\end{equation}
It turns out that (\ref{d}) can be evaluated explicitly:
\begin{equation}\label{appell}
d(r) =\frac{Nr \alpha^{N-1}}{(4 \alpha)^{\frac{5N-8}{2(N+4)}}}f(r)
\end{equation}
where
\begin{widetext}
\begin{equation}
f(r)=(N+1)\mathcal{A}(a_1,1-N,a_2,a_3,a_4,a_5)+(1-2N)\mathcal{A}(a_1,2-N,a_2,a_3,a_4,a_5)+(N-2)\mathcal{A}(a_1,3 -N,a_2,a_3,a_4,a_5),
\end{equation}
\end{widetext}
and where $\mathcal{A}$ is the Appell hypergeometric function F1 \cite{appell} with
\begin{equation}\label{a1}
a_1=\frac{1}{2},
\end{equation}
\begin{equation}\label{a2}
a_2 = \frac{5N-8}{2(N+4)},
\end{equation}
\begin{equation}\label{a3}
a_3 = \frac{3}{2},
\end{equation}
\begin{equation}\label{a4}
a_4 = -\frac{r^2}{\alpha},
\end{equation}
and
\begin{equation}\label{a5}
a_5 = -\frac{(N+4)r^2}{4 \alpha}.
\end{equation}
Whereas one could argue that the choice (\ref{tolman}) is \textit{ad hoc}, the real question is whether or not the choice of (\ref{tolman}) leads to explicit solutions that satisfy the conditions \textbf{1} through \textbf{5}. This is examined below.
\subsection{Boundary Surfaces}

\begin{figure}[ht]
\epsfig{file=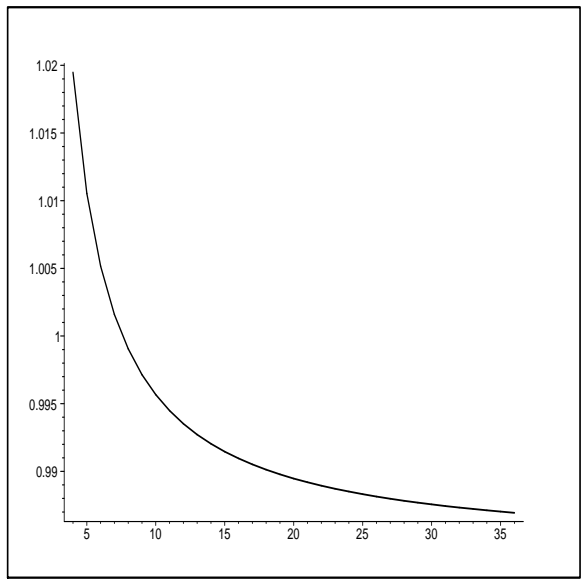,height=3in,width=3in,angle=270}
\caption{\label{figure1} Evolution of $R/R_B$ (ordinate) as a function of $N$ (abscissa) for $\alpha=2$.}
\end{figure}

It is easy to show that boundary surfaces $p(R)=0$ exist only for $N \geq 4$, and we restrict our attention to this range. Moreover it follows that
\begin{equation}\label{boundary}
\frac{R}{R_B} = \frac{9R^2(1+2N)+\alpha}{4NR^2}
\end{equation}
where $R_B$ represents the the Buchdahl limit. For $N \geq 8$ these configurations break the Buchdahl bound (but, of course satisfy the Andr\'{e}asson bound). $R/R_{B}$  is most conveniently found numerically. The results are shown in Figure \ref{figure1}.
\subsection{Interior Structure}
\begin{figure}[ht]
\epsfig{file=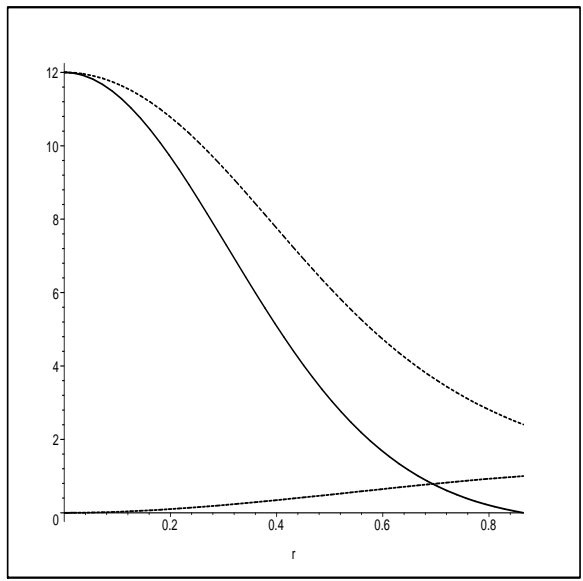,height=3in,width=3in,angle=270 }
\caption{\label{figure2} $p$ (solid), $P$ (dot) and $(P-p)/P$ (bottom) for $N = 16$ and $\alpha=2$.}
\end{figure}

\begin{figure}[ht]
\epsfig{file=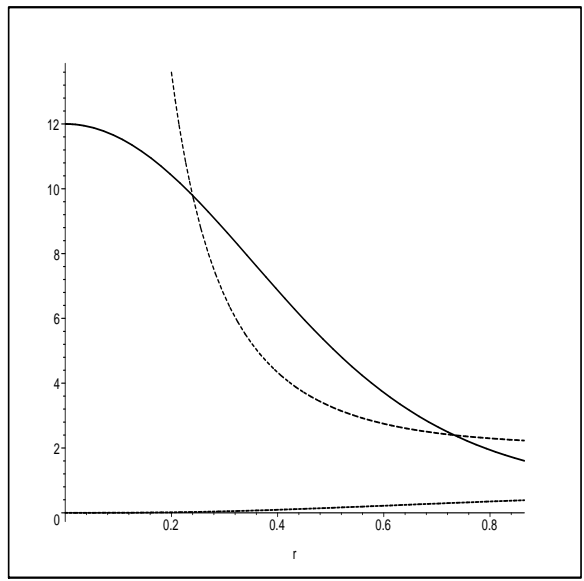,height=3in,width=3in,angle=270 }
\caption{\label{figure3} $\rho$ (solid), $r/m$ (dot) and $m$ (bottom) for $N = 16$ and $\alpha=2$.}
\end{figure}

\begin{figure}[ht]
\epsfig{file=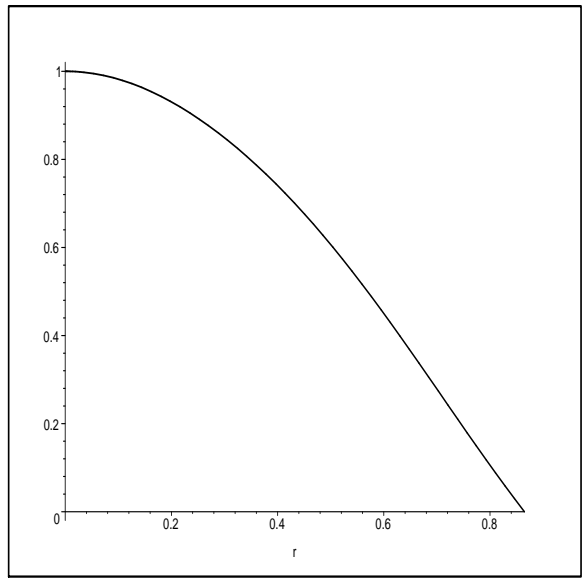,height=3in,width=3in,angle=270 }
\caption{\label{figure4} $p/\rho$ for $N = 16$ and $\alpha=2$.}
\end{figure}

With(\ref{tolman}) and $N\geq4$ the physical quantities $\rho, p$ and $P$ are monotonically decreasing in $r$, and the equation of state becomes stiff at the origin since
\begin{equation}\label{origin}
 \lim_ {r \rightarrow 0}\frac{p}{\rho} \rightarrow 1.
\end{equation}
The evolution of $p$, $P$ and $(P-p)/P$ is shown in Figure \ref{figure2} and the evolution of $\rho$, $r/m$ and $m$ is shown in Figure \ref{figure3} for $N = 16$ and $\alpha = 2$. Figure \ref{figure4} shows $p/\rho$.
\subsection{Internal Trapping}
The evolution of non-radial null geodesics is governed by the potential impact parameter \cite{ishak}
\begin{equation}\label{impact}
\mathcal{B}(r) \equiv \frac{r}{e^{\Phi(r)}}.
\end{equation}
Since, in general,
\begin{equation}\label{impactmax}
p\big | _{\mathcal{B}^{'}=0}=\frac{2}{r^3}\left(r-3m(r)\right),
\end{equation}
and since regularity requires $\mathcal{B}(0)=0$, the solutions described here all exhibit internal photon trapping. An example is shown in Figure {\ref{figure5}
\begin{figure}[ht]
\epsfig{file=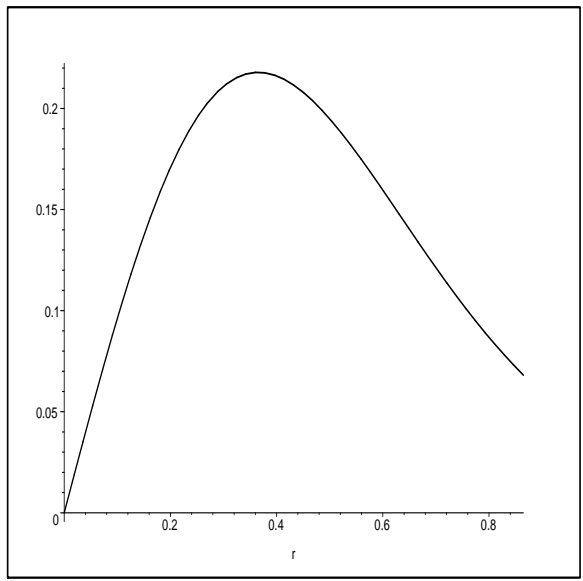,height=3in,width=3in,angle=270 }
\caption{\label{figure5} $\mathcal{B}(r)$ for $N = 16$ and $\alpha=2$.}
\end{figure}
\subsection{Adiabatic Sound Speed?}
It is not uncommon to find
\begin{equation}\label{sound}
v_s^2 \equiv \frac{dp}{d \rho}
\end{equation}
and to refer to $v_s$ as the adiabatic ``speed of sound".
For example, in the case of the familiar Schwarzschild interior solution, $\rho =$ const. $p=P$ ,
use of $v_s$ would suggest an \textit{infinite} adiabatic sound
speed (the incompressible limit). Yet, even in this extreme case,
it could be argued that the adiabatic sound speed is
inappropriate. The distribution $\rho =$ const.  may in fact model
an object with a (contrived) composition variation rendering $v_s$
meaningless as regards the speed of sound \cite{mtw}. Here, from (\ref{dominant}), we have
\begin{equation}\label{sound1}
v_s^2 =3 -2 \frac{dP}{d \rho},
\end{equation}
which need not be monotone. However, numerical analysis shows that $v_s^2$ is monotone decreasing with $r$ for $N \geq 5$. The case $N = 4$ is shown in Figure \ref{figure6}.

\begin{figure}[ht]
\epsfig{file=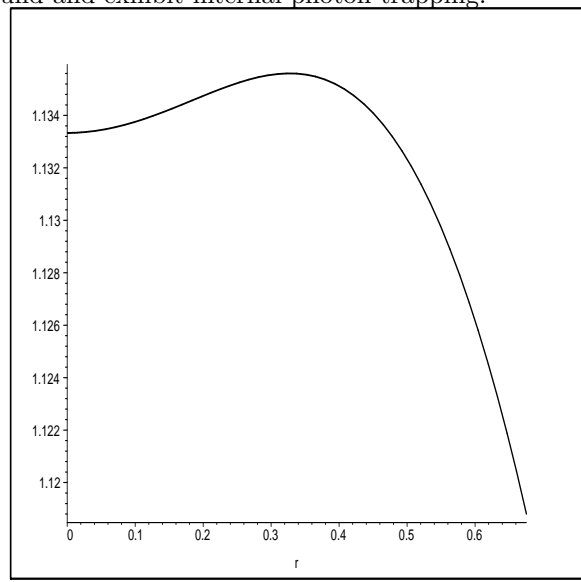,height=3in,width=3in,angle=270 }
\caption{\label{figure6} $v_s^2$ for $N = 4$ and $\alpha=2$. The range in $r$ is $0$ to $R/5$.}
\end{figure}

The reliable calculation of a physically realistic sound speeds requires knowledge of parameters (like chemical concentrations, entropy density and so on) that the model considered here does not provide (see, for example Rahman and Visser in \cite{visser}).
\section{Discussion}
An infinite class of exact explicit static anisotropic spheres that satisfy the conditions \textbf{1} through \textbf{5} have been developed. All of these solutions break the Buchdahl bound and exhibit internal photon trapping.

\textit{Acknowledgments.} This work was supported by a grant from the Natural Sciences and Engineering Research Council of Canada. Portions of this work were made possible by use of \textit{GRTensorII} \cite{grt}.


\begin{thebibliography}{}
\bibitem{footnote} The dominant energy condition has as its motivation the idea that matter should flow along timelike (or null) world lines. The fact that $P$ is not associated with such a flow renders the condition $\rho \leq P$ innocuous.
\bibitem{notation} We use geometrical units and usually designate functional dependence only on the first appearance of a function. We assume that all functions are at least $C^2$.
\end{thebibliography}
\end{document}